\documentclass[aps, prl, showpacs, showkeys, preprint, floatfix, superscriptaddress, nofootinbib, longbibliography]{revtex4-2}

\usepackage{hyperref}
\usepackage{graphicx}
\usepackage{amsmath}
\usepackage{amssymb}
\usepackage{amsthm}
\usepackage{bm}
\usepackage{relsize}

\def\vp{\varphi}
\def\ra{\rightarrow}

\def\half{\textstyle{\frac{1}{2}}}

\def\ra{\rightarrow}

\def\bn{\begin{eqnarray*}}  %takes no eqn numbers
\def\en{\end{eqnarray*}}    %takes no eqn numbers
\def\bq{\begin{eqnarray}}  %takes eqn numbers
\def\eq{\end{eqnarray}}   %takes eqn numbers

\begin{document} 
%%%%%%%%%%%%%%%%%%%%%%%%%%%%%%%%%%%%%%%%%%%%%%%%%%%%%%%%%%%%%%%%%%%%%%%%%%%%%%
%%%%%%%%%%%%%%%%%%%%%%%%%%%%%%%%%%%%%%%%%%%%%%%%%%%%%%%%%%%%%%%%%%%%%%%%%%%%%%
%%%%%%%%%%%%%%%%%%%%%%%%%%%%%%%%%%%%%%%%%%%%%%%%%%%%%%%%%%%%%%%%%%%%%%%%%%%%%%
\title{Scaled Affine Quantization of $\vp^{12}_3$ is Nontrivial}

\author{Riccardo Fantoni}
\email{riccardo.fantoni@posta.istruzione.it}
\affiliation{Universit\`a di Trieste, Dipartimento di Fisica, \\
strada Costiera 11, 34151 Grignano (Trieste), Italy}
%\author{John R. Klauder}
%\email{klauder@ufl.edu}
%\affiliation{Department of Physics and Department of Mathematics \\
%University of Florida,   %P.O. Box 118440\\
%Gainesville, FL 32611-8440}

\date{\today}

\begin{abstract}
We prove through path integral Monte Carlo that the covariant euclidean scalar field 
theory, $\vp^r_n$, where $r$ denotes the power of the interaction term and $n = s + 1$ 
with $s$ the spatial dimension and $1$ adds imaginary time, such that $r = 12, n = 3$ can 
be acceptably quantized using scaled affine quantization and the resulting theory is 
nontrivial, unlike what happens using canonical quantization which finds it trivial. 
\end{abstract}

\maketitle
%%%%%%%%%%%%%%%%%%%%%%%%%%%%%%%%%%%%%%%%%%%%%%%%%%%%%%%%%%%%%%%%%%%%%%%%%%%%%%
\section{Introduction}
%%%%%%%%%%%%%%%%%%%%%%%%%%%%%%%%%%%%%%%%%%%%%%%%%%%%%%%%%%%%%%%%%%%%%%%%%%%%%%
Covariant euclidean scalar field quantization, henceforth denoted $\vp^r_n$, where $r$ 
is the power of the interaction term and $n = s + 1$ with $s$ the spatial 
dimension and $1$ adds imaginary time, such that $r < 2n/(n - 2)$ can be treated by 
canonical quantization (CQ), while models such that $r \geq 2n/(n - 2)$ are trivial 
\cite{Freedman1982,Aizenman1981,Frohlich1982,Siefert2014,Aizenman2021}. However, there 
exists a different approach called affine quantization (AQ) 
\cite{Klauder2000,Klauder2020c} that promotes a different set of classical variables to 
become the basic quantum operators and it allows to correctly quantize such models 
\cite{Fantoni2020,Fantoni2020a,Fantoni2020b,Fantoni2021,Fantoni2021a,Fantoni22b,Fantoni22c,Fantoni22d} which appear then
to be nontrivial. In the present work we show, with the aid of a path integral Monte Carlo 
(MC) analysis, that one of the special cases where $r > 2n/(n - 2)$, specifically the case 
$r = 12, n = 3$, can be acceptably quantized using {\sl scaled} affine quantization. This 
work complements the previous one \cite{Fantoni2020} where the same analysis was carried 
out in the unscaled version. Such unscaled version was later found to have some shortcoming 
like a diverging vacuum expectation value of the field in the continuum limit 
\cite{Fantoni2020b}. In the subsequent work \cite{Fantoni2021} we discovered that a simple 
scaling procedure would cure such a divergence, even if this was done for a complex field. 
It was then not obvious whether the necessary scaling procedure would keep the affinely 
quantized real field theory nontrivial. In the present work we will answer affirmatively to 
this question. The same analysis has been carried out for the $\vp^4_4$ theory in 
\cite{Fantoni22a}.  

%%%%%%%%%%%%%%%%%%%%%%%%%%%%%%%%%%%%%%%%%%%%%%%%%%%%%%%%%%%%%%%%%%%%%%%%%%%%%%
\section{Affine version of the field-theory}
%%%%%%%%%%%%%%%%%%%%%%%%%%%%%%%%%%%%%%%%%%%%%%%%%%%%%%%%%%%%%%%%%%%%%%%%%%%%%%
\label{sec:affine}
Our quantum covariant Euclidean relativistic field theory model has a standard 
Hamiltonian given by,
\bq
H[\pi,\varphi]=\mathop{\mathlarger{\mathlarger{\int}}}\left\{\frac{1}{2}
\left[\pi^2(x)+\sum_{\mu=1}^s\left(\frac{\partial\varphi(x)}{\partial x_\mu}\right)^2 
+m^2\varphi^2(x)\right]+g\varphi^r(x)\right\}\,d^sx,
\eq
where $s$ denotes the number of spatial coordinates and $x_0$ is the time. The 
momentum field $\pi(x)=\partial\phi(x)/\partial x_0$ and the {\sl canonical} action 
is $S=\int H \,dx_0$.

Next, we introduce the affine field $\kappa(x)\equiv\pi(x)\varphi(x)$, with 
$\varphi(x)\neq 0$ and modify the classical Hamiltonian to become 
\cite{Klauder2020,Klauder2020b,Klauder2020c},
\bq
H^\prime[\kappa,\varphi]=\mathop{\mathlarger{\mathlarger{\int}}}\left\{\frac{1}{2}
\left[\kappa(x)\varphi^{-2}(x)\kappa(x)+\sum_{\mu=1}^s\left(\frac{\partial\varphi(x)}
{\partial x_\mu}\right)^2+m^2\varphi^2(x)\right]+g\varphi^r(x)\right\}\,d^sx.
\eq
In an affine {\sl quantization} the operator term 
$\widehat{\kappa}(x)\varphi^{-2}(x)\widehat{\kappa}(x)=\widehat{\pi\mkern 0mu}^2(x)+\hbar^2(3/4)\delta^{2s}(0)\varphi^{-2}(x)$ 
which leads to an extra ``$3/4$'' potential \cite{Gouba2020} term. So that the new 
{\sl affine} action will formally read,
\bq \label{eq:affine-action-old}
S^\prime[\varphi]=\mathop{\mathlarger{\mathlarger{\int}}}\left\{\frac{1}{2}
\left[\sum_{\mu=0}^s\left(\frac{\partial\varphi(x)}{\partial x_\mu}\right)^2 
+m^2\varphi^2(x)\right]+g\varphi^r(x)+\frac{3}{8}\hbar^2\frac{\delta^{2s}(0)}{\varphi^2(x)+
\epsilon}\right\}\,d^nx,
\eq 
where $\epsilon>0$ is a parameter used to regularize the ``$3/4$'' extra term. In 
the $g\to 0$ limit, this model remains different from a free-theory due to 
the new $(3/8)\hbar^2\delta^{2s}(0)/[\phi^2(x)+\epsilon]$ interaction term.

In order to explain the extra ``$3/4$'' potential term we use the fact that 
the operator corresponding to the affine field $\kappa$ will be the {\sl dilation} 
operator 
$\widehat{\kappa}=(\widehat{\pi}\widehat{\varphi}+\widehat{\varphi}\widehat{\pi})/2$ 
where the regularized basic quantum Schr\"odinger operators are given by 
$\widehat{\varphi}(x)=\varphi(x)$ and 
$\widehat{\pi}(x)=-i\hbar\delta_{\varphi(x)}=-i\hbar\delta/\delta\varphi(x)$ so that the 
commutator
$[\widehat{\varphi}(x),\widehat{\pi}(y)]=i\hbar\delta^s(x-y)$, where $\delta^s(x)$ is 
a $s$-dimensional Dirac delta function since 
$\delta_{\varphi(x)}\varphi(y)=\delta^s(x-y)$. Multiplying this by $\widehat{\varphi}$ 
we find 
$[\widehat{\varphi},\widehat{\varphi}\widehat{\pi}]=[\widehat{\varphi},\widehat{\pi}\widehat{\varphi}]=[\widehat{\varphi},\widehat{\kappa}]=i\hbar\delta^{s}\widehat{\varphi}$ 
which is only valid for $\varphi\neq 0$. Then
$\widehat{\kappa}=-i\hbar\{\delta_{\varphi(x)}[\varphi(x)]+\varphi(x)\delta_{\varphi(x)}\}/2=-i\hbar\{\delta^s(0)/2+\varphi(x)\delta_{\varphi(x)}\}$. 
Now, for $\varphi(x)\neq 0$, we will have that affine quantization sends 
$\widehat{\pi\mkern 0mu}^2(x)$ to
\bq \nonumber
\widehat{\kappa}(x)\varphi^{-2}(x)\widehat{\kappa}(x)&=&
-\hbar^2\{\delta^s(0)/2+\varphi(x)\delta_{\varphi(x)}\}\varphi^{-2}(x)
\{\delta^s(0)/2+\varphi(x)\delta_{\varphi(x)}\}\\ \nonumber
&=&\hbar^2(3/4)\delta^{2s}(0)\varphi^{-2}(x)-\hbar^2\delta_{\varphi(x)}^2\\
&=&\hbar^2(3/4)\delta^{2s}(0)\varphi^{-2}(x)+\widehat{\pi\mkern 0mu}^2(x).
\eq

%%%%%%%%%%%%%%%%%%%%%%%%%%%%%%%%%%%%%%%%%%%%%%%%%%%%%%%%%%%%%%%%%%%%%%%%%%%%%%
\section{Lattice formulation of the field theory}
%%%%%%%%%%%%%%%%%%%%%%%%%%%%%%%%%%%%%%%%%%%%%%%%%%%%%%%%%%%%%%%%%%%%%%%%%%%%%%
The theory considers a real scalar field $\vp$ taking the value $\vp(x)$ on each site of 
a periodic, hypercubic, $n$-dimensional lattice of lattice spacing $a$, our ultraviolet 
cutoff, and periodicity $L=Na$. The affine action for the field is then 
$S'=\int H'\,dx_0$, with $x_0=ct$ where $c$ is the speed of light constant and $t$ is 
imaginary time, and $H'$ is the Hamiltonian. The lattice formulation of the AQ 
field theory used in Eq. (\ref{eq:affine-action-old}) is
\bq \label{eq:affine-action} \nonumber
S'[\vp]/a^{n}&\approx&\half\left\{\sum_{x,\mu}a^{-2}[\vp(x)-\vp(x+e_\mu)]^2 
+m^2\sum_{x}\vp(x)^2\right\}+\sum_{x}g\,\vp(x)^r\\
&&+{\textstyle\frac{3}{8}\sum_{x}}\hbar^2{\displaystyle\frac{a^{-2s}}{\vp(x)^2+\epsilon}},
\eq
where $e_\mu$ is a vector of length $a$ in the $+\mu$ direction and the factor $a^{-2s}$ 
in the effective interaction term due to the AQ procedure, namely 
$\frac{3}{8}\hbar^2\delta^{2s}(0)/(\vp(x)^2+\epsilon)$, stems from the 
discretization of the Dirac delta function, and diverges in the continuum limit.

In order to solve such divergence \cite{Fantoni22a} we apply the scaling 
$\vp\ra a^{-s/2}\vp, g\ra a^{(r-2)s/2} g, \epsilon\ra a^{-s}\epsilon$ to the action 
(\ref{eq:affine-action}) above, in order to suppress the $a^{-2s}$ factor, which diverges 
in the continuum limit, at the price of having a field of dimensions 
$a^{1/2}$. As a consequence of our scaling, the action becomes
\bq \label{eq:scaled-affine-action} \nonumber
S'[\vp]/a^{n-s}&\approx&\half\left\{\sum_{x,\mu}a^{-2}[\vp(x)-\vp(x+e_\mu)]^2 
+m^2\sum_{x}\vp(x)^2\right\}+\sum_{x}g\,\vp(x)^r\\
&&+{\textstyle\frac{3}{8}\sum_{x}}\hbar^2{\displaystyle\frac{1}{\vp(x)^2+\epsilon}},
\eq

In this work we are interested in reaching the 
continuum limit by taking $Na$ fixed and letting $N\to\infty$ at fixed volume $L^s$ and 
absolute temperature $T=1/k_BL$ with $k_B$ the Boltzmann's constant.

%%%%%%%%%%%%%%%%%%%%%%%%%%%%%%%%%%%%%%%%%%%%%%%%%%%%%%%%%%%%%%%%%%%%%%%%%%%%%%
\section{MC results}
%%%%%%%%%%%%%%%%%%%%%%%%%%%%%%%%%%%%%%%%%%%%%%%%%%%%%%%%%%%%%%%%%%%%%%%%%%%%%%
We performed the path integral MC \cite{Metropolis,Kalos-Whitlock,Ceperley1995} 
calculation for the AQ field theory previously done in \cite{Fantoni2020} for the case 
$r =  12, n = 3$ using now the scaling 
$\vp\ra a^{-s/2}\vp, g\ra a^{5s} g, \epsilon\ra a^{-s}\epsilon$, which brings to 
using the lattice formulation for the action of Eq. (\ref{eq:scaled-affine-action}).
In particular we calculated the renormalized coupling constant $g_R$ and mass $m_R$ 
defined in Eqs. (4.3) and (4.5) of \cite{Fantoni2020} respectively.

\begin{figure}[htbp]
\begin{center}
\includegraphics[width=9cm]{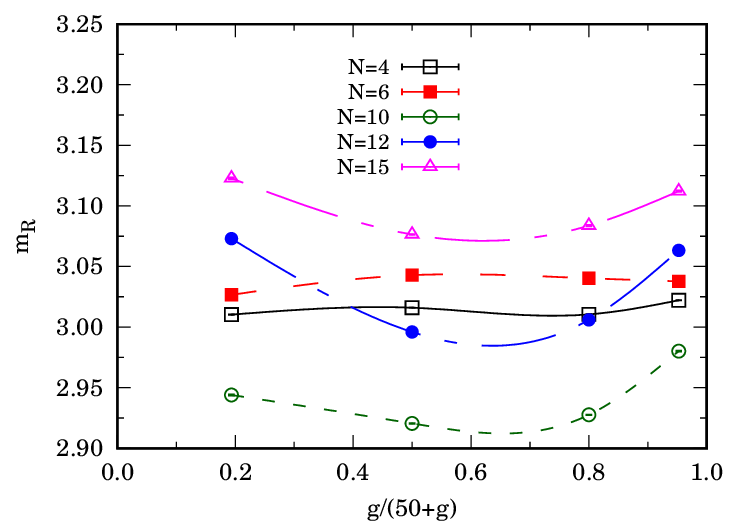}\\
\includegraphics[width=9cm]{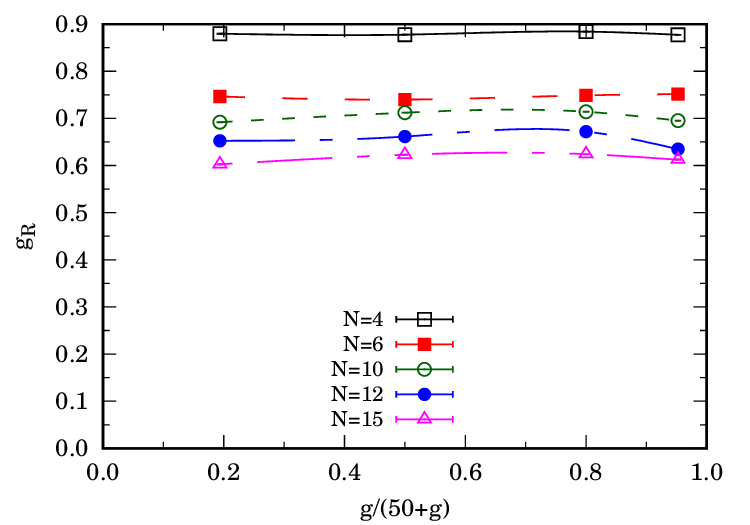}\\
\includegraphics[width=9cm]{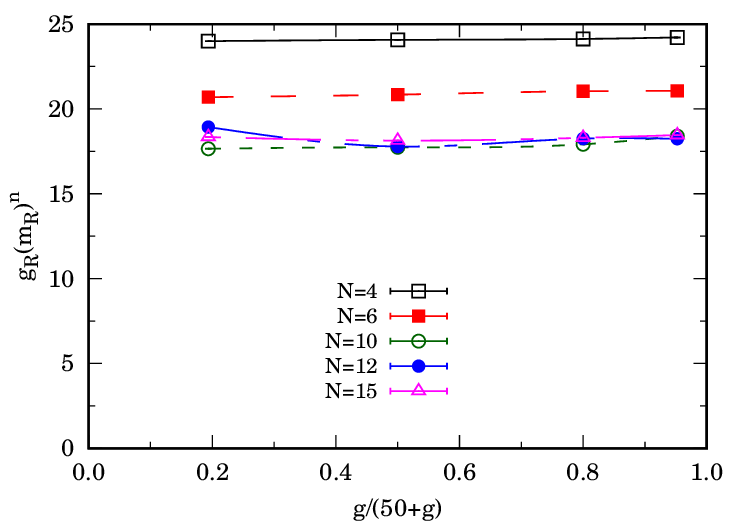}
\end{center}
\caption{(color online) We show the renormalized mass $m_R\approx 3$ (top panel), the 
renormalized coupling constants $g_R$ (central panel), and $g_Rm_R^n$ (bottom panel) 
for various values of the bare coupling constant $g$ at decreasing 
values of the lattice spacing $a=1/N$ ($N\to\infty$ continuum limit) for the 
{\sl scaled affine} $\vp^{12}_3$ covariant euclidean scalar field theory described by 
the action in Eq. (\ref{eq:scaled-affine-action}) for $r = 12,  n = 3$. The statistical 
errors were in all cases smaller than the symbols used. The lines connecting the 
simulation points are just a guide for the eye.}
\label{fig:123}
\end{figure}

For each $N$ and $g$, we adjusted the bare mass $m$ in such a way to maintain the renormalized mass approximately constant $m_R\approx 3$ to within a few percent (in all cases less than $5\%$) \cite{Freedman1982}. Differently from our previous study \cite{Fantoni2020} with the unscaled version of the affine 
field theory we did not need to choose complex $m$ in order to fulfill this constraint. 
Moreover the needed $m$ turned out to be independent from $g$. Then we measured the 
renormalized coupling constant $g_R$ defined in \cite{Fantoni2020,Fantoni2020a} for 
various values of the bare coupling constant $g$ at a given small value of the lattice 
spacing $a=1/N$ (this corresponds to choosing an absolute temperature $k_BT=1$ and a 
fixed volume $L^2=1$). The results are shown in Fig. \ref{fig:123} for the scaled affine action (\ref{eq:scaled-affine-action}) in natural units $c=\hbar=k_B=1$ and 
$\epsilon=10^{-10}$ (the results are independent from the regularization parameter as 
long as this is chosen sufficiently small). The fixed renormalized mass constraint 
was not easy to implement since for each $N$ and $g$ we had to run the 
simulation several times with different values of the bare mass $m$ in order to 
determine the value which would satisfy the constraint $m_R\approx 3$. 

In our simulations we always used $3\times 10^7$ MC steps (which took about one week of 
computer time for the $N=15$ case). We estimated that it took roughly $10-50\%$ of each 
run in order to reach equilibrium from the arbitrarily chosen initial field 
configuration, for each set of parameters. We needed longer equilibration times for 
bigger $N$. Our MC simulations use the Metropolis algorithm 
\citep{Kalos-Whitlock,Metropolis} to calculate the required $N^n$ multidimensional 
integrals. The simulation is started from the initial condition $\phi=0$. One MC step 
consisted in a random displacement of each one of the $N^n$ components of $\phi$ as 
follows: $\phi\rightarrow\phi+(\eta-1/2)\delta$, where $\eta$ is a uniform pseudo random 
number in $[0,1]$ and $\delta$ is the amplitude of the displacement. Each one of these 
$N^n$ moves is accepted if $\exp(-\Delta S')>\eta$ where $\Delta S'$ is the change in the 
action due to the move (this can be efficiently calculated considering how the kinetic 
part and the potential part change by the displacement of a single component of $\phi$)
and rejected otherwise. The amplitude $\delta$ is then chosen in such a way to have 
acceptance ratios as close as possible to $1/2$ and is kept constant during the 
evolution of the simulation.

These results should be compared with the results of Figure 1 of 
\cite{Fantoni2020} where the same calculation was done for the canonical version of 
the field theory. As we can see from our present Figure, contrary to Figure 1 of 
\cite{Fantoni2020}, the renormalized coupling constant $g_R(m_R)^3$ of the scaled 
affine version remains far from zero in the continuum limit when the ultraviolet cutoff 
is removed ($Na=1$ and $N\to\infty$) for all values of the bare coupling constant $g$. 
Here, unlike in the canonical version used in \cite{Fantoni2020}, the diminishing space 
between higher $N$ curves is a pointer toward a non-free ultimate behavior as  
$N\to\infty$ at fixed volume. Moreover as one can see the $N=15$ results for the 
renormalized coupling fall above the ones for $N=12$. 

During our simulations we kept under control also the vacuum expectation value of the 
field which in all cases was found to vanish in agreement with the fact that the 
symmetry $\vp\to-\vp$ is preserved. We can then say that the scaled system is profoundly 
different from the unscaled one previously treated in \cite{Fantoni2020} where a 
diverging value of the expectation value of the field was found as a result of the broken 
symmetry. This is ultimately due to the fact that the scaling procedure avoids a 
diverging width of the infinite repulsive barrier at $\vp=0$ in the continuum limit and 
this makes possible the crossing of $\vp=0$ by the random walk.

%%%%%%%%%%%%%%%%%%%%%%%%%%%%%%%%%%%%%%%%%%%%%%%%%%%%%%%%%%%%%%%%%%%%%%%%%%%%%%
\section{Conclusions}
%%%%%%%%%%%%%%%%%%%%%%%%%%%%%%%%%%%%%%%%%%%%%%%%%%%%%%%%%%%%%%%%%%%%%%%%%%%%%%
In conclusion we performed a path integral Monte Carlo study of the properties (mass 
and coupling constant) of the renormalized covariant euclidean scalar field theory 
$\vp^{12}_3$ quantized through {\sl scaled} affine quantization. As already pointed out 
in \cite{Fantoni2021} where the complex field was allowed to rotate around the potential 
barrier at $\vp=0$, therefore producing a vanishing field expectation value, we here 
observe that due to the used scaling on the real field, its vacuum expectation value and 
the two-point function are well defined in the continuum limit and not diverging like 
what we observed in \cite{Fantoni2020} without the scaling. More 
importantly, we are still able to show that, unlike what happens for the theory quantized 
through canonical quantization, the renormalized coupling constant $g_R(m_R)^3$ does 
not tend to vanish in the continuum limit, when the ultraviolet cutoff is removed at 
fixed volume. This is a strong indication that affine quantization is indeed able to 
render renormalizable classical field theories which would be otherwise 
nonrenormalizable when treated with canonical quantization because of asymptotic freedom.

%%%%%%%%%%%%%%%%%%%%%%%%%%%%%%%%%%%%%%%%%%%%%%%%%%%%%%%%%%%%%%%%%%%%%%%%%%%%%%
%\bibliographystyle{}
\bibliography{123}

%%%%%%%%%%%%%%%%%%%%%%%%%%%%%%%%%%%%%%%%%%%%%%%%%%%%%%%%%%%%%%%%%%%%%%%%%%%%%%

%%%%%%%%%%%%%%%%%%%%%%%%%%%%%%%%%%%%%%%%%%%%%%%%%%%%%%%%%%%%%%%%%%%%%%%%%%%%%%
%%%%%%%%%%%%%%%%%%%%%%%%%%%%%%%%%%%%%%%%%%%%%%%%%%%%%%%%%%%%%%%%%%%%%%%%%%%%%%
%%%%%%%%%%%%%%%%%%%%%%%%%%%%%%%%%%%%%%%%%%%%%%%%%%%%%%%%%%%%%%%%%%%%%%%%%%%%%%
\end{document}